\documentclass[%
 reprint, prx,
 amsmath,amssymb,
 aps, superscriptaddress]{revtex4-2}

\usepackage{graphicx}
\usepackage{dcolumn}
\usepackage{bm}

\usepackage{amsmath}

\usepackage{graphicx}
\usepackage{dcolumn}
\usepackage{bm}
\usepackage{xcolor}
\usepackage{comment}
\usepackage{subcaption}
\usepackage{soul}
\usepackage[normalem]{ulem}
\usepackage{esint}

\newcommand{\todo}[1]{\textcolor{red}{[#1]}}

\newcommand{\V}[1]{\boldsymbol{ #1}}

\newcommand{\Vh}[1]{\hat{\boldsymbol{ #1}}}

\newcommand{\del}{\V{\nabla}}
\newcommand{\laplacian}{\nabla^2}

\newcommand{\pd}[2]{\frac{\partial #1}{\partial #2}}
\newcommand{\td}[2]{\frac{d #1}{d #2}}
\newcommand{\md}[2]{\frac{D #1}{D #2}}

\newcommand{\eqnref}[1]{eqn.~\ref{eqn:#1}}
\newcommand{\Eqnref}[1]{Eqn.~\ref{eqn:#1}}
\newcommand{\figref}[1]{Fig.~\ref{fig:#1}}

\newcommand{\secref}[1]{\S\ref{sec:#1}}

\begin{document}


\title{From Dye to Dynamics: Quantitative Measurements of Fluid Flows from Vortex Geometry}

\author{Dustin Kleckner}
\email{dkleckner@ucmerced.edu}
\affiliation{Physics Department, University of California Merced}

\author{Kevin Collins}
\affiliation{Physics Department, University of California Merced}

\author{Cole Cooper}
\affiliation{Physics Department, University of California Merced}

\date{\today}

\begin{abstract}
Although vorticity and passive tracers (such as dye) are both transported by fluid flows, there are subtle differences in how this transport happens.
This indicates that dye cannot be used to track vortex motion in experiments, even though it has often been used for this purpose.
In this manuscript we show that under proper circumstances dye can be used to accurately track the center-lines of vortex filaments.
Moreover, we show that a number of important fluid quantities, such as energy and helicity, can be computed in terms of the shape of the vortex center-line.
As a result, dye-tracked vortices are a valuable tool for measuring the properties of fluid motion in a variety of experimental situations.
Moreover, we discuss techniques by which could be used to extract more detailed vortex information in future work.
\end{abstract}

\maketitle

\section{Introduction}
\label{sec:intro}
The traditional representation of a fluid flow is in terms of the velocity field, $\V u(\V r, t)$.
Alternatively, fluid flows can be expressed in terms of the curl of the velocity, $\V \omega = \del \times \V u$, known as vorticity.
For incompressible flows ($\del \cdot \V u = 0$), vorticity provides a \emph{complete} description of the flow.
Moreover, in high Reynolds number flows we commonly find compact filaments of intense vorticity, referred to as a `vortex'.
Examples of compact vortex filaments are found in weather systems (both terrestrially and on other planets \cite{balmeDustDevilsEarth2006, rotunnoFluidDynamicsTornadoes2013, adrianiClustersCyclonesEncircling2018}), turbulence \cite{sheIntermittentVortexStructures1990, douadyDirectObservationIntermittency1991, jimenezStructureIntenseVorticity1993a} and flows around aerodynamic surfaces \cite{crowStabilityTheoryPair1970a, breitsamterWakeVortexCharacteristics2011}.
As a result, there is a long history of using mathematically simple vortex models to explain complex fluid phenomena without resorting to expensive direct numerical simulations of the Navier-Stokes equations.

More recently, experimental techniques have emerged for directly tracking vortices in experiment, as an alternative to traditional velocity field measurements \cite{klecknerCreationDynamicsKnotted2013, mckeownCascadeLeadingEmergence2018, mckeownTurbulenceGenerationIterative2020, scheelerCompleteMeasurementHelicity2017, tapiasilvaHighspeedTwocolorScanning2024}.

If vortices are the primary structures of interest in a fluid flow, directly tracking them offers a significant advantage over trying to reconstruct them from a flow field. 
Although in principle the vorticity can be computed as a curl of a measured flow field, taking numerical derivatives introduces significant noise \cite{luffExperimentalUncertaintiesAssociated1999, chartrandNumericalDifferentiationNoisy2011}. 
Moreover typical velocity field reconstruction techniques, such as particle image velocimetry (PIV) \cite{westerweelParticleImageVelocimetry2013, raffelParticleImageVelocimetry2018}, have a resolution that is typically 1-2 orders of magnitude lower than the raw images recorded of an experiment.

Tracking vortices can be accomplished by either injecting dye into the cores of vortices as they are created \cite{limInstabilityReconnectionHeadon1992, mckeownCascadeLeadingEmergence2018, mckeownTurbulenceGenerationIterative2020, scheelerCompleteMeasurementHelicity2017}, or seeding the fluid with small gas bubbles which get trapped in the vortex cores \cite{douadyDirectObservationIntermittency1991, klecknerCreationDynamicsKnotted2013}.
With modern imaging systems, high resolution volumetric data at video rates or faster can be used to track vortex lines at a resolution roughly equal to the resolution of the image \cite{mckeownTurbulenceGenerationIterative2020, silvaHighspeedTwocolorScanning2023}. This leads to two important questions:
(1) do the tracers remain accurate indicators of regions of compact vorticity as the flow evolves, and (2) can \emph{quantitative} information be obtained from the reconstructed vortex lines?
In this manuscript we make a first-principles argument that the geometry of a vortex \emph{center-line} can be tracked using active or passive tracers in vortex-dominated flows.
Moreover, we show that important quantities like energy, helicity, and enstrophy can be computed from the vortex centerline geometry if some reasonable assumptions are made.
Finally, we discuss some potential methods for improving and expanding these measurements in future work.

\section{Transport of Vorticity, Dye, and Bubbles}
\label{sec:transport}

The time evolution of vorticity is obtained by taking the curl of the Navier-Stokes equation.
If we assume an incompressible Newtonian fluid with dynamic viscosity $\mu$ and no body forces, we obtain what is known as the vorticity transport equation \cite{majdaVorticityIncompressibleFlow2001, batchelorIntroductionFluidDynamics2010}:
\begin{align}
    \md{\V \omega}{t} &= \V \omega \cdot \del\V u + \nu  \laplacian \V \omega \label{eqn:vorticity-transport}
    \\\label{eqn:dwbn} \Rightarrow \pd{\V \omega}{t} &= \V \omega \cdot \del \V u - \V u \cdot \del \V \omega + \nu \laplacian \V \omega
\end{align}
where $\md{\V \omega}{t} \equiv \pd{\V \omega}{t} + \V u \cdot \del \V \omega$ is the material derivative, which describes the rate of change of vorticity when following an individual fluid parcel, and $\nu$ is the kinematic viscosity.
The first term on the right hand side of this equation describes how the lines of vorticity are concentrated or reoriented by the gradient of the flow field.
As we will show below, one important effect of this term is to concentrate vorticity when a vortex is stretched along its length.
Note that this term is absent in two dimensional incompressible flows, meaning that the vorticity density cannot be concentrated or dissipated by stretching.
This effect  has important consequences for the behavior of 2D vs.~3D turbulence \cite{ouelletteTurbulenceTwoDimensions2012}.
The last term in \eqnref{vorticity-transport} describes the effects of viscosity, which causes the vorticity to diffuse, thickening a vortex filament in time.

\subsection{Passive Tracers}
\label{sec:passive-tracers}

The material derivative of the density of a passive tracer (such as fluorescent dye) is given by the advection-diffusion equation \cite{batchelorIntroductionFluidDynamics2010}:
\begin{align}
    \pd{C}{t} + &\del \cdot \left(C \V u - D \del C\right) = 0\\
    \label{eqn:pt} \Rightarrow \md{C}{t} &= -C \del \cdot \V u + D \laplacian C
\end{align}
where $D$ is the diffusion coefficient of the passive tracer, and $C$ is its concentration.
In an incompressible flow (i.e., $\del \cdot \V u = 0$) this becomes:
\begin{align}
    \md{C}{t} &= D \laplacian C \label{eqn:advection-diffusion}
    \\\label{eqn:dp} \Rightarrow \pd{C}{t} &= -\V u \cdot \del C + D \laplacian C
\end{align}
There are two important differences when comparing \eqnref{advection-diffusion} to \eqnref{vorticity-transport}.
Firstly, the directional derivative term is absent from the dye transport.
This prohibits the concentration of dye by the advection of the fluid.
The second difference is that the coefficient diffusive term is given by the dye diffusivity instead of the viscosity. 
The typical diffusivity of commonly used florescent fluid tracers is on the order of $D \lesssim 5 \times 10^{-10}$ m$^2$/s 
(for example: Rhodamine derivatives or Fluorescein dyes \cite{culbertsonDiffusionCoefficientMeasurements2002, gendronDiffusionCoefficientsSeveral2008}).   
In contrast, the kinematic viscosity of fluids is typically $\nu \gtrsim 10^{-7}$ m$^2$/s; for example pure water $\nu \sim 10^{-6}$ m$^2$/s.
Thus, in most experimental situations the diffusion of vorticity is four or more orders of magnitude faster than that of dye; in practice the dye diffusion is almost entirely negligible for moderate to high Reynolds number flows where vortex-based methods are appropriate.

The differences between these transport equations indicate that dye \emph{density} cannot be used to track vorticity density, even if it is seeded properly at some initial time.
(An analytical example of this for a straight, twisted vortex is discussed in  \secref{twisted}.)
Does this mean that dye cannot be used to track the shape of a vortex filament as it evolves?
As we will show below, concentrated vortex lines behave like material lines in the flow; meaning that the center-line of a vortex tube will follow a line of dye injected there at the beginning of an experiment.

To see why this is the case, consider a thin line of dye placed in a fluid following some path $\V c(\ell, t)$, where $\ell$ is the path-length coordinate at $t=0$.
The position of this dye line is a `material line', which is to say it follows the fluid parcels:
\begin{align}
\pd{\V c(\ell, t)}{t} &= \V u(\V c(\ell, t))
\end{align}
Now consider how the tangent vector of this line evolves:
\begin{align}
    \Vh t(\ell, t) = \pd{\V c}{\ell} \Rightarrow \md{\Vh t}{t} = \pd{\V u}{\ell} = \Vh t \cdot \del \V u
\end{align}
where we note that, in general, $\Vh t$ is only unit length at $t=0$ because the material line will be stretched in time.
Interestingly, the transport of the tangent recovers the tilting/stretching term seen in \eqnref{vorticity-transport}; ($\Vh t$ points in the same direction as $\V \omega$ on the vortex centerline).
As a result, the center-line of a vortex filament is itself a material line; if the core of a vortex is seeded with dye at some initial time, the dye will stay in the vortex core even as the vortex line evolves into a distorted shape.

In order for a dye line to be a faithful tracer of the vortex filament, we must assume that the effects of diffusion (i.e.~viscosity) do not displace the vortex filament center-line with respect to a material line.
This requires that (1) the vortex is a compact bundle of vorticity in an otherwise irrotational flow, and (2) the vortex does not contact any other region of compact vorticity, like other vortices or walls.
In the latter case a vortex reconnection will result; typically this occurs when regions of opposite vorticity annihilate by rapidly approaching and diffusing into each other \cite{kidaVortexReconnection1994}.
In this case, the region of space where the reconnection occurred will have little or no vorticity remaining, while some dye would be left behind.
This can be mitigated by using buoyant tracers, as discussed in the next section.
Finally, we note that the requirement that the background flow have no vorticity could be relaxed if the background flow had constant and/or slowly varying vorticity with comparison to the vortex thickness.
In this case, the diffusion of the vorticity would still be roughly uniform in all directions, in which case a material line would still be an accurate tracer of its center-line.

\subsection{Buoyant Tracers}
An alternative to passive tracers is to use buoyant particles or bubbles which are attracted to the vortex cores.
Due to centripetal pressure, vortex filaments form low pressure centers which have been used to track vortices in both classical fluids and superfluids \cite{vargaDynamicsDensityQuantized2018, klecknerCreationDynamicsKnotted2013,caridiHeliumfilledSoapBubbles2017}.
Another option is to use a gas solution near saturation, in which case the low pressure centers can cause bubbles to nucleate out of the liquid phase \cite{douadyDirectObservationIntermittency1991}.

Once bubbles have been attracted to the vortex core, they should remain there as long as the filament remains compact enough that the pressure gradient exceeds gravitational forces.
We can estimate a capture radius for bubbles by finding the distance at which centripital acceleration is equal to gravity:
\begin{align}
    \frac{u^2_{\phi}}{s} &= \frac{\Gamma^2}{4 \pi^2 s^3} > g\\
    \Rightarrow \Gamma &> 2 \pi \sqrt{g s^3}
\end{align}
where $s$ is the radial distance out from the center of a vortex with circulation $\Gamma$. We assuming that $s > a$, where $a$ is the vortex filament core size (see next section).
For example, assuming a capture radius of $s=5$ mm and $g = 9.8$ m/s$^2$ results in a minimum circulation of $\Gamma = 7000$ mm$^2$/s.
In general, this technique works better for larger systems; for small systems such a high circulation would result in fast vortex motion that would be difficult to resolve.

\section{Analytical Solution to a Twisted Core Vortex}
\label{sec:twisted}

Although the discussion in the previous makes a general theoretical argument that compact lines of dye should follow vortex center-lines, it is instructive to consider what is happening in the region near a vortex core.
Below we consider exact solutions to the Navier-Stokes equation which extends the familiar Lamb-Oseen and Burger's vortex \cite{rottViscousCoreLine1958, saffmanVortexDynamics1993}.
We present solutions of twisted core vortices both in isolation and in stretching flows.
In both cases we illustrate the difference in transport of vorticity and dye.

\subsection{Isolated Solution}

Consider the vorticity and flow field of a `twisted core' vortex, which in cylindrical coordinates ($s$, $\phi$, and $z$) can be written as:
\begin{align}
    \label{eqn:tc} \V \omega &= \underbrace{\frac{\Gamma}{2\pi a^2} e^{-\frac{s^2}{2a^2}}}_{\omega_0(s)} \left(\Vh z + s \tau \Vh \phi\right)\\
    \Rightarrow \V u &= \frac{\Gamma}{2 \pi}\left[ \left(\frac{1 - e^{-\frac{s^2}{2a^2}}}{s}\right)\Vh \phi + \left(\tau e^{-\frac{s^2}{2a^2}} \right)\Vh z\right]
    \end{align}
where $a$ is a Gaussian core radius, and $\tau$ is twist angle per unit length.
Note that this is equivalent to a Lamb-Oseen vortex if $\tau = 0$ and $a = \sqrt{2\nu t}$ \cite{saffmanVortexDynamics1993}.
Similarly, this is equivalent to a Batchelor vortex with $\tau = (\sqrt{2} q a)^{-1}$ \cite{batchelorAxialFlowTrailing1964}.

Assuming we have a uniform external flow field, the evolution of the dynamics of this system is governed by the diffusive term, $\nu \laplacian \omega$, only, resulting in: 
\begin{align}
    \pd{a}{t} &= \frac{\nu}{a}& \Rightarrow a(t) &= a_0 \sqrt{1 + t/t_c} \label{eqn:dyn-a}\\
    \label{eqn:dtau} \pd{\tau}{t} &= -\frac{2 \nu \tau}{a^2} &\Rightarrow \tau(t) &= \frac{\tau_0}{1 + t/t_c}\\
    \pd{\Gamma}{t} &= 0
\end{align}
where $t_c = a_0^2 / 2\nu$ is the characteristic viscous diffusion timescale for the core.

The time evolution of the core radius, $a$, can be understood simply as diffusion of vorticity, where the diffusion coefficient is given by the kinematic viscosity.
We note that the total circulation, $\Gamma$, is not affected by this diffusion.
Similarly, the decay of twist can be understood in terms of vorticity diffusion; in a sense the $\Vh \phi$ component of the vorticity is pointing in opposite directions on either side of the vortex filament.
Diffusion causes annihilation of this component, which reduces the net twist in the vortex over time.
(This can equivalently be understood in terms of helicity dissipation, as described in \secref{helicity}.)

\subsection{Evolution of Dye Density}
Suppose our vortex is initially seeded with dye having a Gaussian density profile, i.e.:
\begin{align}
    C &= \frac{\lambda}{2\pi b^2} e^{-\frac{s^2}{2b^2}}
\end{align}
where $\lambda$ is the linear dye density.
The time evolution of the dye (per \eqnref{advection-diffusion}) is given by:
\begin{align}    
    \pd{b}{t} &= \frac{D}{b}& \Rightarrow b(t) &= \sqrt{b_{0}^2 + 2 D t}\\
    \pd{\lambda}{t} &= 0
\end{align}
If the dye radius is initially equal to the vortex core radius, i.e.~$b_0 = a_0$, the dye will continue to track the vorticity \emph{density} if and only if the diffusion rate of the dye matches the kinematic viscosity, $D = \nu$.
As discussed in \secref{passive-tracers}, this is not achievable with typical experimental conditions.
As a result, dye cannot be used to determine a vortex's overall extent, in most cases it is preferable to start with a compact dye core, $b \ll a$, and then use it to indicate the evolution of the vortex \emph{center-line}.

\subsection{Stretching Flows}

\label{sec:stretching}

Let us now consider the effect of a stretching or compressing flow along the vortex axis, given by an incompressible external flow field:
\begin{align}
    \V u_{ext} &= \alpha \left(2 z\, \Vh z - s\, \Vh s\right)
\end{align}
where $\alpha$ is the strain rate, defined in the same manner as the Burgers Vortex.

If we solve for the time variation of vortex core size and twist in this flow, we obtain additional terms from \eqnref{vorticity-transport} \cite{delbendeStretchingEffectsThreedimensional2002, gibbonDynamicallyStretchedVortices1999}:
\begin{align}
    \pd{a}{t} &= \frac{\nu}{a} - \alpha a
        &\Rightarrow a^2 = \frac{\nu}{\alpha} + \left(a_0^2 - \frac{\nu}{\alpha}\right) e^{-2\alpha t} \label{eqn:da_dt_stretch}\\
    \pd{\tau}{t}& = -\left(\frac{2\nu}{a^2} + 2\alpha\right)\tau
        &\Rightarrow \tau a^2 = \tau_0 a_0^2\, e^{-4\alpha t} \label{eqn:dtau_dt_stretch}\\
    \pd{\Gamma}{t} &= 0
\end{align}
The core size relaxes to the steady value $a = \sqrt{\nu/\alpha}$, which recovers the Burgers vortex, an exact steady solution of the Navier--Stokes equations.
The twist, however, has no corresponding steady state: once the core has reached its steady size, $\tau$ decays as $e^{-4\alpha t}$.

We can derive a similar solution for a dye line with a Gaussian distribution, resulting in:
\begin{align}
    \pd{b}{t} &= \frac{D}{b} - \alpha b\\
    \pd{\lambda}{t} &= -2 \alpha \lambda
\end{align}
Again we observe that the dye line radius evolves in a similar manner to the vortex core, apart from the difference in diffusion speed.
However, unlike the vortex case -- which conserves total circulation, $\Gamma$ -- the linear dye density, $\lambda$, is not conserved.

This difference can be understood by noting that positive $\alpha$ is a stretching flow. 
If we consider a section of the line with length $\ell$, the change in length of this segment will be given $\dot \ell = 2 \alpha \ell$.
As a result, we note that the total amount of dye in this segment is given by $\ell \lambda$, which \emph{is} conserved under this stretching flow.
We note that the net twist angle, $\tau \ell$ is conserved in a similar manner if one neglects the viscous dissipation term.
On the other hand, circulation is not affected by stretching the vortex filament, indeed the vorticity \emph{density} actually increases as core size is decreased.
(For incompressible flows the \emph{volume} of a vortex filament section is conserved if we neglect viscosity.  As a result, we expect that $\pd{(a^2 \ell)}{t} = 0 \Rightarrow \pd{a}{t} = -\frac{a}{2 \ell} \pd{\ell}{t} = -\alpha a$.)

Finally we note that one could also consider the addition of an overall background flow to the stretching term.  
However, as noted in the previous section, both dye and vorticity are advected by the overall flow, therefore a non-stretching flow would only advect both in an identical manner.
In other words, we only expect deviation in the behavior of dye vs.~vorticity with respect to stretching flows.

\subsection{Conservation of Circulation}
\label{sec:circulation}
Conservation of circulation for a vortex filament is usually understood in terms of the Kelvin circulation theorem, which can be derived for an inviscid, barotropic fluid subject only to conservative forces \cite{thomsonVortexMotion1868, saffmanVortexDynamics1993}
\begin{align}
    \label{eqn:if}\md{\V u}{t} &= -\frac{\del p}{\rho} + \del \Phi\\
    \Rightarrow \md{\Gamma}{t} &= \md{}{t}\left[\oint_{\V c} \V u \cdot d\V \ell\right] \\
    &= \oint_C \md{\V u}{t} \cdot d\V \ell + \oint_c \V u \cdot (d\V \ell \cdot \del) \V u = 0
\end{align}
Where $\md{\Gamma}{t}$ indicates we are measuring the circulation as the vorticity flux through a loop which is transported by the flow in time.

In the previous section, we showed that circulation is conserved even for flows with viscosity. This can be understood by adding a viscous term to \eqnref{if}:
\begin{align}
    \md{\V u}{t} &= \nu \laplacian \V u-\frac{\del p}{\rho} + \del \Phi\\
    \Rightarrow \md{\Gamma}{t} &= \oint_C (\nu \laplacian \V u) \cdot d\V \ell\\
    &= \nu\bigg[\underbrace{\oint_c \del (\del \cdot \V u) \cdot d\V \ell}_0 -  \oint_0 \del \times \V \omega \cdot d\V \ell\bigg]
\end{align}
where we have made use of the fact that $\laplacian \V u = \del (\del \cdot \V u) - \del \times (\del \times \V u)$.

Let us consider the total circulation for the twisted core vortex above.
The curl of the vorticity for the twisted core vortex is given by:
\begin{align}
    \del \times \V \omega &= \omega_0(s)\left[\frac{s}{a^2}\Vh \phi + \left(2-\frac{s^2}{a^2}\right) \tau \Vh z\right]
\end{align}
with $\omega_0(s) = \Gamma e^{-s^2 /2 a^2}/(2\pi a^2)$ as defined in \eqnref{tc}.
Suppose we measure the circulation by drawing a loop of radius $r$ around the vortex core (oriented in the $d\V \ell \propto \Vh \phi$ direction), in which case we obtain:
\begin{align}
    \md{\Gamma(r)}{t} = -2\pi \nu \frac{r^2 }{a^2}\omega_0(r) = -\frac{\Gamma}{t_c} \left(\frac{r^2}{2a^2} e^{-r^2 /2 a^2}\right)
\end{align}
If the loop is infinitely large, $r \rightarrow \infty$, we note that $\Gamma(r) \rightarrow \Gamma$ and so $\md{\Gamma(r)}{t} \rightarrow 0$.
In other words, the circulation is always conserved if the loop is big enough; this is because viscosity causes diffusion of dissipation, but does not annihilate it.

However, we note that $\omega_0(s)$ also rapidly decays; for $r = 5a$ we obtain $\md{\Gamma(r)}{t} < 5 \times 10^{-5} \frac{\Gamma}{t_c}$.
In other words, provided we measure the circulation using a material line that is sufficiently outside the core -- and will remain so as it diffuses -- we expect that the circulation will be conserved to good approximation.
Similarly, stretching or shearing flows will not change the circulation, as they transport the material line used to measure $\Gamma$ so that it stays centered around the vortex filament.

\begin{figure}
    \includegraphics[width=\columnwidth]{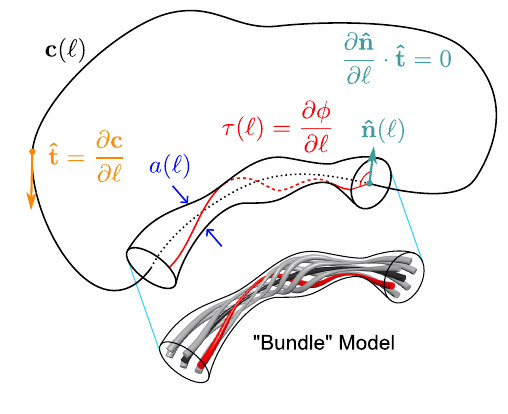} 
    \caption{\label{fig:bundle}
        A schematic of the `bundle' model used to derive physical quantities.
        In such a model, finite core size and twist are included, as opposed to approximating the vortex filament as infinitesimally thin and ignoring internal twist.
        The bundle can be visualized as a division of our vortex into sub-filaments, which may twist around the center-line.
    }
\end{figure}

\section{Derivation of Physical Quantities from Centerline Shape}
\label{sec:centerline}

For an unbounded incompressible flow of constant density, the velocity, $\V u(\V r)$ induced by a general vorticity field, $\V \omega(\V r)$ can be obtained via the Biot-Savart law:
\begin{align}
    \V u(\V r) &=\frac{1}{4 \pi}\int \frac {\V \omega(\V r') \times \V R'}{R'^3} dV' 
\end{align}
where $\V R' \equiv \V r - \V r'$ is the displacement vector.

Consider a flow system where the vorticity is confined to closed, compact bundles (\figref{bundle}), each parameterized by its arc length $\ell_i$ with centerline curve $\V c(\V \ell_i)$, circulation $\Gamma_i$, core size $a(\ell_i)$, and twist $\tau(\ell_i)$. 
If we assume the cores are relatively compact (i.e. $\kappa a \ll 1$, where $\kappa$ is the centerline curvature), and neglect the twist component, we can approximate the flow field as:
\begin{align}
    \label{eqn:biot-savart-line}
    \V u_{cl}(\V r) \cong \sum_i\frac{\Gamma_i}{4\pi} \oint_{\V c} g_u\left(R/a\right) \frac{\Vh t \times \V R}{R^3} d\ell_i
\end{align}
where $\Vh t = \partial \V c / \partial\ell_i$ is the tangent of each curve, and $\V R(\ell_i) = \V r - \V c(\ell_i)$ the displacement vector. 
(Note that we have omitted the subscript $i$ for compactness, but it is implicitly included for all quantities referring to an individual vortex bundle.)
The integral is regularized by introducing a short-range cutoff function, $g_u(R / a)$, which vanishes on the vortex centerline and approaches unity outside the core region, where the field coincides with that of an infinitely thin line vortex.
Common choices for $g_u(R / a)$ include a hard cutoff \cite{saffmanVortexDynamics1993, branlardWindTurbineAerodynamics2017} or simple polynomial functions like \todo{XX, ref}.
This regularization captures the finite thickness of the vortex bundle and is essential to avoid divergences of the velocity, energy, and related quantities on the vortex core.
Note that each of these quantities should have a slightly different regularization function, indicated by subscripts.
In all cases we assume $g(0) = 0$ and $g\left(\frac{R}{a} \gg 1\right) = 1$, so the function has no effect when fully outside the core.

We define the twist of the bundle relative to a parallel transport (PT) frame, also known as a Bishop \cite{bishopThereMoreOne1975} or rotation minimizing frame \cite{wangComputationRotationMinimizing2008}. 
This is defined by a normal vector $\Vh n_i(\ell_i)$ such that $\Vh n_i \perp \Vh t_i$ and $\pd{\Vh n}{\ell} \cdot \Vh t = 0$.
Let $\phi(\ell_i)$ denote the angle between $\Vh n$ and a second normal vector that points from the centerline toward a given vorticity integral line (e.g., the red line in \figref{bundle}); the twist is then $\tau = \pd{\phi}{\ell_i}$.
The additional velocity field induced by the twist is approximately:
\begin{align}
    \V u_{tw} \cong \frac{\tau \omega_0(q)}{a^2} \Vh t,
\end{align}
where $q$ is the radial distance to the nearest vortex filament, and $\omega_0$ is the Gaussian vorticity density profile as previously defined in \secref{twisted}.

This definition of twist differs from that obtained using a Frenet-Serret frame.
However, parallel transport framing is typically preferable for measuring twist because it remains well defined and stable even where the curvature becomes small or vanishes, thereby avoiding the frame singularities and rapid rotations that occur with the Frenet-Serret frame \cite{moffattHelicityCalugareanuInvariant1992}.
Moreover, it can be theoretically and experimentally demonstrated that viscous relaxation will produce a state with $\tau=0$ as measured with respect to a parallel transport framing \cite{scheelerCompleteMeasurementHelicity2017}, making it a physically informed choice (see \secref{framing}).

It is also useful to define a vector potential, $\V A$, such that $\V u = \del \times \V A$ and $\V \omega = -\nabla^2 \V A$.
(This is analogous to the magnetic vector potential, with the substitutions $\V B \rightarrow \V u$ and $\mu_0 \V J \rightarrow \V \omega$.)
It then follows that:
\begin{align}
    \V A_{cl} &= \frac{1}{4\pi} \int \frac{\V \omega(\V r')}{R'} dV'\\
    &\cong \sum_i \frac{\Gamma_i}{4\pi}\int g_A\left(\frac {R'}{a}\right) \frac{\Vh t'}{R'} d\ell_i' \label{eqn:Acl}
\end{align}
where $g_A$ is a modified regularizing function for the vector potential.
We can produce an expression for $g_A$ by comparing the curl of the \eqnref{Acl} to \eqnref{biot-savart-line}, resulting in $g_u(\Delta) = g_A(\Delta) - \Delta \pd{g_A(\Delta)}{\Delta}$.

Using these forms, it is possible to compute many quantities of interest in fluids in terms of the geometry of the vortex lines, by integrating over $\ell_i$ and $q$, and making use of the fact that $\int 2\pi q\ (\V \omega \cdot \Vh t)\ dq = \Gamma$.
Note that these forms are correct only in the weak curvature limit $\kappa a \gg 1$.
Although it should be possible to derive corrections to these quantities, in most cases they should appear at order $(\kappa a)^2$ or higher.
As a result, the precise form of the regularizing functions typically has a small effect on the computed quantities, and mathematically convenient forms can be chosen in practice.

\subsection{Energy}
\label{sec:energy}

The energy of a flow field is given by:
\begin{align}
    E &= \frac{\rho}{2}\int u^2 dV\\
    &= \frac{\rho}{2} \left[\int \left(\V A \cdot \V \omega \right) dV + \oint\left( \V u \times \V A\right) \cdot d\V a\right]
\end{align}
where $\rho$ is the fluid density which is assumed to be constant.
In the second line we have made use of the fact that $\V u \cdot \V u = \del \cdot (\V u \times \V A) + \V A \cdot \V \omega$, and the integrals are taken over some finite volume and its boundary.
In the limit that the volume is all of space, the surface integral vanishes and the volume integral in terms of vortex bundles results in:
\begin{align}
    \label{eqn:energy-cl}
    E_{cl} &\cong \frac{\rho}{2} \sum_{i, j} \frac{\Gamma_i \Gamma_j}{4\pi} \oiint_{\V c_i, \V c_j} g_E\left(\frac{R}{a_i}\right) \frac{\Vh t_i \cdot \Vh t_j}{R_{ij}}\, d\ell_i\, d\ell_j
\end{align}
where here $\V R_{ij} = \V c(\ell_i) - \V c(\ell_j)$, $g_E$ is a regularization function for the energy, $\Vh t_i \equiv \Vh t(\ell_i)$, $\Vh t_j\equiv \Vh t(\ell_j)$, and $a_i \equiv a(\ell_i)$.
If we approximate $\V A$ as constant over the bundle cross-section, we obtain $g_E = g_A$, however this will slightly overestimate the energy as we expect $\V \omega \cdot \V A$ to decrease as we move away from the core.

If we assume individual vortex bundles are not in contact, we can eliminate the regularizing function for $i \neq j$.
Furthermore, provided the center-line is smooth on the scale of $a$, the form of $g$ is not particularly important, apart from an overall correction factor that depends on the core profile (see \secref{coreint}).
In this case, a simple hard cutoff model can be used \cite{denglerSelfInductanceWire2016}, resulting in \cite{klecknerCreationDynamicsKnotted2013}:
\begin{widetext}
\begin{align}
     \label{eqn:Ehc}
     E_{cl} \cong \frac{\rho}{2}\bigg[\sum_i& \Gamma^2_i \left( \frac{1}{4\pi}\oiint_{\V c_i, \V c_{i'}} \Theta (R_{i i'} - a) \frac{\Vh t_i \cdot \Vh t_{i'}}{R_{i i'}}\, d\ell_i \, d\ell_{i'} + Y L_i\right)+\sum_{i\neq j} \frac{\Gamma_i \Gamma_j}{4\pi} \oiint_{\V c_i, \V c_j}  \frac{\Vh t_i \cdot \Vh t_j}{R_{ij}}\,d\ell_i\, d\ell_j\bigg]
\end{align}
\end{widetext}
where $\Theta$ is the unit step function, $Y$ is a core correction parameter that depends on the core profile \cite{donnellyVortexRingsClassical2009}, and $L_i$ is the total length of each vortex bundle.
This form ignores an additional -- and usually much weaker -- contribution from twist, which will produce flow velocity only in the vortex core.

\subsection{Energy Dissipation and Enstrophy}

For a real vortex, we expect the core to thicken in time due to diffusion, i.e.~$a \rightarrow a(\ell_i, t)$.
We can compute the rate of change of energy due only to the change of core thickness as: 
\begin{align}
    \label{eqn:energy-cl-diss}
    \mathcal{E}_{a} &= -\left[ \pd{E_{cl}}{t}\right]_{a(\ell) \rightarrow a(\ell, t)} \cong \frac{\rho}{2} \sum_i \frac{\Gamma_i^2}{2 \pi} \oint  \pd{a}{t}\frac{1}{a}\ d\ell_i 
\end{align}
Where we have assumed that $a$ is small enough that \mbox{$\Vh t(\ell_i)\cdot \Vh t(\ell_i + a) \cong 1$}, and made use of the fact that the derivative of the step function is a delta-function at $\ell_i' = \ell_i \pm a$.

If we assume a Gaussian core, which thickens like \eqnref{dyn-a}, this reduces to: $\mathcal{E}_{a,g} \cong \frac{\rho\nu}{2} \sum_i \frac{\Gamma_i^2}{2 \pi} \oint  a^{-2}\ d\ell_i$.
By comparison, the enstrophy of the flow in our bundle mode is given by:
\begin{align}
    \Omega = \int |\omega|^2 dV &\cong \sum_i \oint_c d\ell_i \int \omega^2_0(q) 2\pi q\ dq\\
    &\cong -\sum_i C_a \Gamma_i^2 \oint_c a^{-2} \ d\ell_i \label{eqn:enstrophy},
\end{align}
where $q$ is the coordinate perpendicular to the bundle centerline, and $C_a$ is a constant that depends on the core model.
In particular, $C_a = 1/4\pi$ for a Gaussian core, in which case we observe that $\mathcal{E}_{a, g} = \rho \nu \Omega_{g}$.
This is an important result: the total energy dissipation in a flow should be proportional to the enstrophy. This suggests that, for non-contacting vortex bundles, the \emph{only} source of energy loss is core thickening.
In other words, the motion of the centerline must otherwise conserve energy when computed with a fixed core size.
This has important consequences, for example in constraining the geometry of reconnecting vortex lines \cite{laingConservationWritheHelicity2015, scheelerHelicityConservationFlow2014}, or in the growth of colliding vortex rings \cite{limInstabilityReconnectionHeadon1992, mckeownCascadeLeadingEmergence2018}.

\begin{figure}
  {
    \centering
    \includegraphics[width=\linewidth]{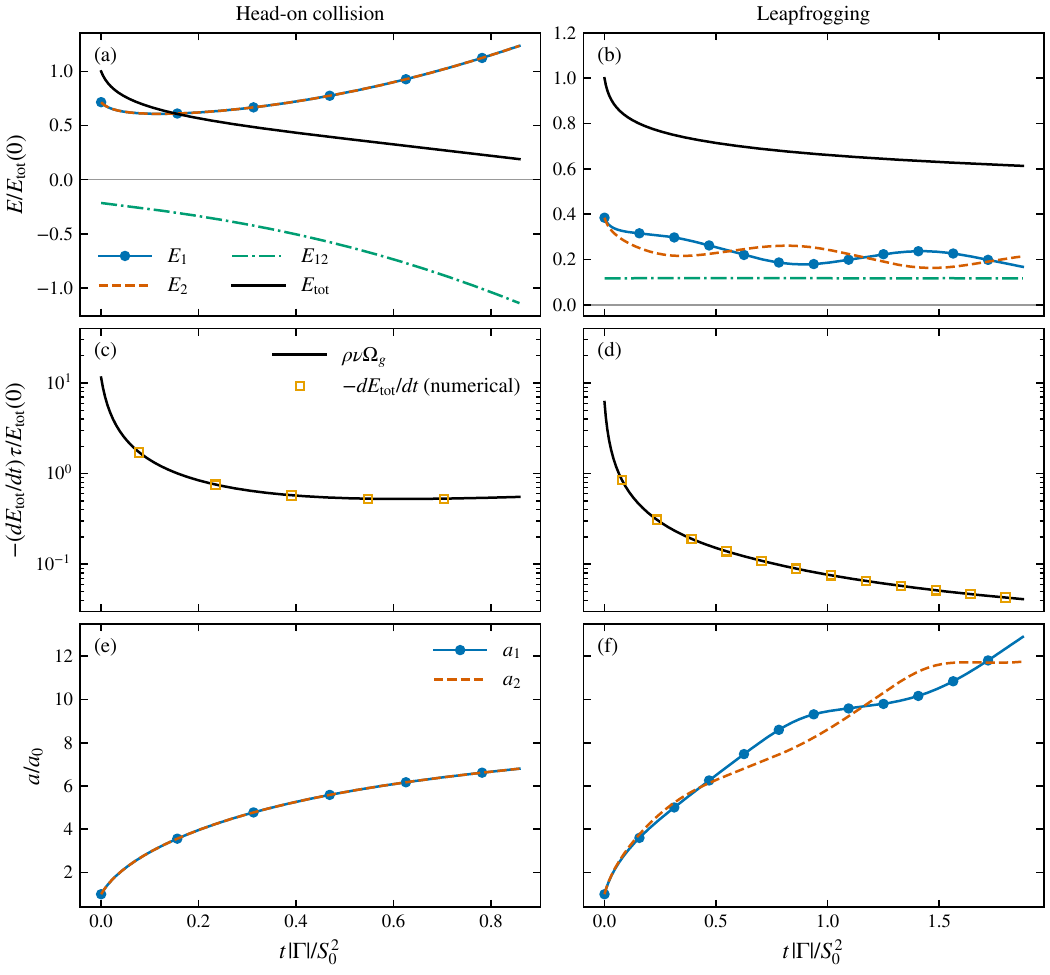}
  }
  \caption{
    (c, d) Energy dissipation rate from the analytical expression \(\rho \nu \Omega_g\) (solid line) compared with the numerical time derivative \(-dE_{tot}/dt\) of the simulated total energy (squares), normalized by \(E_{tot}(0)/\tau\).
    (e, f) Core radii \(a_1\) and \(a_2\) normalized by the initial core radius \(a_0\); the decay of the dissipation rate, which scales as \(a^{-2}\), follows from this core growth. Time is normalized by \(\tau = S_0^2/|\Gamma|\).
    The collision simulation ends at \(t/\tau \approx 0.86\), when the axial separation of the rings falls below the sum of their core radii.
}
  \label{fig:axisymmetric-rings}
\end{figure}

As a demonstration of our energy dissipation formulation, we constructed an axisymmetric model of colliding and leapfrogging vortex rings.
The self-induced velocity and energy of the ring in each system can be computed from known results \cite{donnellyVortexRingsClassical2009, saffmanVortexDynamics1993}, and the cross terms between neighboring vortices can be analytically computed from the Biot-Savart law (\eqnref{biot-savart-line}) and the centerline derived energy (\eqnref{energy-cl}) by assuming the vortices are non-overlapping (i.e.~$g_{u/E} \rightarrow 1$).
(See \secref{int-circ} for model details.)
We note near-perfect agreement between the numerically computed energy dissipation and the enstrophy derived term, \eqnref{energy-cl-diss}, even as the rings stretch or compress one another (\figref{axisymmetric-rings}). 

\subsection{Helicity}
\label{sec:helicity}
The helicity is given by:
\begin{align}
    H &= \int\left(\V u \cdot \V \omega\right) dV\\
    &\cong \sum_{i, j} \frac{\Gamma_i \Gamma_j}{4\pi} \oiint g\left(\frac{R_{ij}}{a_i}\right)
    \frac{\Vh t_i \times \Vh t_j\cdot \V R_{ij}}{R_{ij}^3} d\ell_i d\ell_j
\end{align}
we note that for smooth curves this integral does not diverge at short range.
Thus the cutoff function can be removed with negligible effect on the result, and the integral becomes equivalent to either the Gauss linking integral (for $i \neq j$) or the equation for writhe (if $i = j$).
The helicity can thus be written as \cite{moffattDegreeKnottednessTangled1969, scheelerCompleteMeasurementHelicity2017}:
\begin{align}
    H = \sum_i \Gamma_i^2\left[Wr_i + \frac{1}{2\pi}\oint_{c_i} \tau d\ell\right] + \sum_{i \neq j} \Gamma_i \Gamma_j Lk_{ij} 
    \label{eqn:helicity-topo}
\end{align}
where we have also added a term corresponding to internal twist in each vortex.
This can be seen either by direct computation or by noting the topological nature of helicity: it essentially measures the total linking of all vortex filaments in a fluid flow \cite{moffattDegreeKnottednessTangled1969}.
Twisting the core creates linking between the filaments of an individual bundle, creating helicity.
Note that the writhe also measures linking between filaments in a bundle, except that it is caused by the net twist resulting from a non-planar centerline.

This presentation of the problem implies a framing with which to measure the twist with respect to.
As previously mentioned, we make use of a parallel transport framing to avoid issues associated with more common frames like the Frenet-Serret.
Consider a curvilinear cylindrical-like coordinate system constructed with respect to this framing around the centerline \cite{germanoEffectTorsionHelical1982,mercierEquilibriumStabilityToroidal1964}:
\begin{align}
    \V r(\ell_i, q, \theta) = \V c(\ell_i) + q\left[\cos \phi\ \Vh e_1 +\sin \phi\ \Vh e_2 \right]
\end{align}
where $\Vh e_1 \perp \Vh t$ is an arbitrary unit vector and $\vec e_2 = \Vh t \times \Vh e_1$.
In this coordinate system, the vorticity of a twisted filament is given by a generalization of \eqnref{tc}:
\begin{align}
    \label{eqn:omega-lqt}\V \omega(\ell_i, q, \phi) &= \omega_0(q)\left[\Vh t + q\tau\ \Vh \phi\right] \\
    \Rightarrow \del \times \V \omega& = \left(2 \tau \omega_0 + \tau q \omega_0'\right)\Vh t-\omega_0'\ \Vh \phi
\end{align}
where we note that the core size and twist may be a function of $\ell_i$.

The dissipation rate of helicity is then given by \cite{moffattDegreeKnottednessTangled1969}:
\begin{align}
    \td{H}{t} &= -2\nu \int \V \omega \cdot (\del \times \V \omega) dV\\
    &\cong -4 \nu \sum_i \oint_c \tau d\ell_i \int 2\pi q \ \omega_0^2(q) dq\\
    \label{eqn:dH-twist}&\cong -4 \nu C_a \sum_i\Gamma_i^2 \oint \frac{\tau}{a^2} d\ell_i
\end{align}

This recovers the previously observed result that only the helicity attributable to local bundle twist is dissipated by viscosity \cite{scheelerCompleteMeasurementHelicity2017}.
If we assume the local twist density evolves in the same manner as the analytical solution for an infinite straight line (\eqnref{dtau}) we would obtain:
\begin{align}
\td{H}{t} = -\frac{\nu}{\pi} \sum_i \Gamma^2_i \oint \frac{\tau}{a^2} d\ell_i
\end{align}
This agrees with \eqnref{dH-twist} for a Gaussian core.
Thus we can regard the helicity dissipation as due to the local untwisting of the core by viscous diffusion of vorticity.
We note, however, that the evolution of twist in a vortex bundle can be quite complex: it's distribution can be changed by stretching and converted to/from writhe via centerline distortions \cite{kamienLocalWrithingDynamics1998, wolgemuthTwirlingWhirlingViscous2000}.
However, both of these effects conserve total helicity as a consequence of the C\u{a}lug\u{a}reanu–White–Fuller theorem \cite{moffattHelicityCalugareanuInvariant1992,calugareanuClassesDisotopieNoeuds1961, fullerWrithingNumberSpace1971, whiteSelfLinkingGaussIntegral1969}, and so the dissipation is caused only by the local action of viscosity.

\section{Measurement of Physical Quantities from Experimental Data}
As we have shown above, vortex tracking using dyed cores can provide direct measurement of dynamically evolving vortex center-line geometry.
Additionally, the total circulation of vortex lines can be measured a low resolution reconstruction of the flow field, i.e.~by computing $\oint \V u \cdot d\V \ell$ for any loop enclosing a vortex line.
Moreover, this circulation is conserved in time for non-overlapping vortices (\secref{circulation}), meaning that it does not need to be continually measured over the course of an experiment.
Recent work has shown that it is possible to simultaneously track dye injected into vortex cores and randomly distributed neutrally buoyant tracers, both in 3D and at video rates or higher \cite{silvaHighspeedTwocolorScanning2023}.
This enables an in-situ measurement of circulation with simultaneous high resolution imaging of the vortex centerline shape.

By contrast, the vortex core size -- which does evolve in time -- can not be directly determined without resolved measurements of the velocity in the core region.
In practice this is unlikely to be experimentally feasible if one is also attempting to resolve the large scale geometry of the vortex lines.
(Indeed, relaxing the requirement that the core be resolved is one of the motivating factors behind dye tracking methods.)
As a result, one often needs to resort to approximating the core as fixed size, or perhaps assuming it diffuses in the same manner as a Lamb-Oseen vortex.
This has different effects on the measurement of different quantities: helicity is essentially independent of core details (\eqnref{helicity-topo}), but enstrophy is critical dependent on the core thickness (\eqnref{enstrophy}).
Energy is dependent on core size, but only logarithmically; in this case a reasonable estimate of the core size may be sufficient.

Alternatively, one could attempt to use more sophisticated methods to infer the evolving core size.
One approach would be to model the core size via an advection-diffusion scheme along the lines of \eqnref{da_dt_stretch}.
This would only require knowledge of the viscosity and local stretching rate, which could be obtained by placing tracer particles in the vortex core or modulating the center-line dye density \cite{naitohExperimentalStudyAxial2002, scheelerCompleteMeasurementHelicity2017, caridiHeliumfilledSoapBubbles2017}.
A similar method has previously been used to compute the twist of the vortex core in terms of measured axial flow, even when core scale velocity details cannot be resolved directly \cite{scheelerCompleteMeasurementHelicity2017}.
An even simpler approach would be to treat the core size as uniform along each vortex, $a(\ell_i) \rightarrow a_i$, and then compute an overall stretch rate from the change in total vortex length, $L_i$:
\begin{align}
    \pd{a_i}{t} &= \frac{\nu}{a_i} - \frac{1}{2L_i} \pd{L_i}{t}
\end{align}
To our knowledge, neither approach has previously been used in analyzing experimental data.
Such a technique may be particularly interesting for the study of vortices in turbulent flows, where enstrophy production is a critical mechanism \cite{meneveauLagrangianDynamicsModels2011, moffattStretchedVorticesSinews1994}.
We note, however, that enstrophy can also be understood as a measure of vortex stretching, which can be measured directly from the total vortex length.

\section{Conclusion}
In this manuscript we have shown that dye can act as an experimentally useful tracker of vortex center-lines.
Although there are differences in the advection of vorticity and dye density, provided they are contained in sufficiently compact tubes their center-lines should evolve identically.
Such vortex tracking has traditionally been used to understand qualitative features of high Reynolds number flows, which tend to be dominated by compact vortex tubes.
However, by making 3D measurements of the vortex geometry one can do far more: it is possible to make \emph{quantitative} measurements of energy, helicity, and vortex stretching.
Although this has been done to a limited extent in past experimental work, it was based on a number of simplifying assumptions.
The results presented here offer a theoretical basis for such measurements, and suggest new avenues quantitative fluid measurements from vortex geometry in future work.

\appendix

\section{Energy Correction Term}
\label{sec:coreint}

In \eqnref{Ehc} we state that a hard cutoff can be used for the energy correction provided we include an additional factor, $Y$, which depends on core model.
To justify this statement, consider a circular vortex ring of radius $S$, in which case the energy integral can be explicitly integrated, resulting in:
\begin{align}
    E_{cl} &= \frac{1}{2} \rho S\Gamma^2\left[\ln\frac{4 S}{a} - 2 + 2 \pi Y\right] + \mathcal{O}\left(\frac{S}{a}\right)^2
\end{align}
For a Gaussian core we can derive $Y = 0.0472$ by comparing to a known result for a circular vortex ring \cite{donnellyVortexRingsClassical2009} (see \eqnref{Ering}).

It furthermore follows that \emph{any} core cutoff function is equivalent to a hard cutoff when computing the energy to order $\left(\tfrac{a}{S}\right)^2 = (\kappa a)^2$, where here $\kappa = S^{-1}$ is the constant curvature.
To see this, consider a cutoff function $g_E(\kappa a)$ for which $g_E(\kappa a \gg 1) = 1$, which should always be true for the regularizing function.
In this case:
\begin{align}
    g_E(x) &= \int_0^x \Theta(x - y) g_E'(y) dy\\
    \Rightarrow E_{cl} &\cong \frac{1}{2} \rho R\Gamma^2\bigg[\underbrace{\left(\int_0^\infty g_E'(y) \ln \frac{4 R}{y a}\, dy\right)}_{=\left(\ln\frac{4 R}{a} - \int_0^\infty g_E'(y) \ln y\, dy\right)} - 2\bigg]\\
    \Rightarrow Y &= -\frac{1}{2\pi}\int_0^\infty g_E'(y) \ln y\, dy
\end{align}
In general, however, the energy regularizing function is not analytically known, and it is easier to derive the correction factor, $Y$, from known results for vortex rings.

\section{Interaction of Circular Vortices}
\label{sec:int-circ}
As an illustration of the vortex dynamics, we construct a simple model of interacting circular vortices on a common symmetry axis.
In this case, each vortex (index $i$) is characterized by a radius, $S_i$, axial position, $z_i$, circulation, $\Gamma_i$, and core size $a_i$.
The self-induced velocity and energy can be computed from known results \cite{donnellyVortexRingsClassical2009, saffmanVortexDynamics1993}
\begin{align}
    v_{i, z} &= \frac{\Gamma_i}{4\pi S_i}\left[\ln \frac{8 S_i}{\sqrt{2}\, a_i} - \beta\right]\\
    E_i &= \frac 1 2 \rho \Gamma_i^2 S_i \left[\ln \frac{8 S_i}{\sqrt{2}\, a_i} - \alpha\right] \label{eqn:Ering}
\end{align}
where $\alpha$ and $\beta$ are dimensionless factors that depend on the core model; for a viscous Gaussian core $\alpha = 2.05$ and $\beta = 0.558$ \cite{donnellyVortexRingsClassical2009}.
We also note that $a$ as defined in the manuscript differs by a factor of $\sqrt 2$ from that in \cite{donnellyVortexRingsClassical2009}.

In addition to the self-induced velocity, the rings also advect one another.
This can be computed in terms of the velocity induced by ring $i$ on ring $j$ \cite{simpsonSimpleAnalyticExpressions2001}:
\begin{align}
    v_{ij, s} &= C_3 \frac{z_{ij}}{S_j} \left[(S_i^2 + S_j^2 + z_{ij}^2 ) E(k^2) - C_1^2 K(k^2) \right]\\
    v_{ij, z} &= C_3 \left[(S_i^2 - S_j^2 - z_{ij}^2 ) E(k^2) + C_1^2 K(k^2) \right]\\
    C_1^2 &\equiv S_i^2 + S_j^2 + z_{ij}^2 - 2 S_i S_j\\
    C_2^2 &\equiv S_i^2 + S_j^2 + z_{ij}^2 + 2 S_i S_j\\
    C_3 &= \frac{\Gamma_i}{2 \pi C_1^2 C_2 }\\
    k^2 &\equiv 1-\frac{C_1^2}{C_2^2}
\end{align}
where $z_{ij} \equiv z_j - z_i$, and $K(k)$ and $E(k)$ are complete elliptic integrals of the first and second kind.
The resulting equations of motion are:
\begin{align}
    \pd{S_i}{t} &= \sum_{i \neq j} v_{ij, s}\\
    \pd{z_i}{t} &= v_{i, z}+ \sum_{i \neq j} v_{ij, z}\\
    \pd{a_i}{t} &= \frac{\nu}{a} - \frac{a}{2 S_i} \pd{S_i}{t}
\end{align}
where $\Gamma_i$ is constant, as we are assuming the vortex bundles do not overlap.
To obtain the time derivative of core size, we assume the solution in \secref{stretching} applies locally to the vortex bundle, where the stretching rate is computed from changing vortex ring radius.

The above equations were implemented numerically to generate \figref{axisymmetric-rings}.
Lengths are scaled by the initial ring radius, $S_0$, and time by the advective time scale $\tau = S_0^2/|\Gamma|$.
Both rings start with the same radius, core size $a_0/S_0 = 0.005$, and circulation magnitude $|\Gamma|$, at Reynolds number $\mathrm{Re} = |\Gamma|/\nu = 1000$, and are initially separated along the axis by $\Delta z_0/S_0 = 0.25$.
We consider two configurations: a head-on collision, in which the rings have opposite circulation ($\Gamma_1 = -\Gamma_2$) and travel toward one another, and leapfrogging, in which the circulations are equal ($\Gamma_1 = \Gamma_2$) and the rings pass through one another in turn.
The equations were integrated with an adaptive fourth/fifth-order Runge--Kutta scheme (relative tolerance $10^{-9}$) up to $t/\tau = 1.875$.
The collision was stopped at $t/\tau \approx 0.86$, when the axial separation of the rings fell below the sum of their core sizes and the vortex bundles could no longer be treated as non-overlapping.
In both configurations, the analytical dissipation rate agrees with the numerical time derivative of the total energy [\figref{axisymmetric-rings}(c, d)], and its decay follows the growth of the cores [\figref{axisymmetric-rings}(e, f)].

\section{Bundle Framing and Calculation of Twist}
\label{sec:framing}
Consider a vortex field line, $\V f$, which is part of a vortex bundle, but which is offset from the centerline:
\begin{align}
    \V f&= \V c + q\Vh n=  \V c + q\left[\cos \theta\ \Vh e_1 + \sin \theta\ \Vh e_2\right],
\end{align}
where $\theta$, $q$, and $\Vh n$, and $c$ are all assumed to be functions of path length coordinate, $\ell$.
$\Vh e_1(\ell)$, and $\Vh e_2(\ell)$ form an orthonormal frame together with the tangent vector, $\Vh t$.
The azimuthal component of the field line direction is given by:
\begin{align}
    \label{eqn:f-az}\Rightarrow \pd{\V f}{\ell} \cdot \Vh \theta &= q \pd{\theta}{\ell} +  \left(\cos \theta\ \pd{\Vh e_1}{\ell} + \sin \theta\ \pd{\Vh e_2}{\ell} \right) \cdot \Vh \theta
\end{align}
By comparison with \eqnref{omega-lqt}, we observe that $\tau = \frac{1}{q}\td{\V f}{\ell} \cdot \hat \theta$, which is identified with the angular rate at which any given field line rotates around the centerline.
The second term in \eqnref{f-az} arises from the twist of the framing itself around the centerline; if we eliminate this contribution it results in the intuitive form of $\tau = \td{\theta}{\ell}$.
By evaluating $\Vh \theta$, one can show this second term can be eliminated provided:
\begin{align}
    \label{eqn:de1-pt}\td{\Vh e_1}{\ell} \cdot \Vh e_2 = 0 \Rightarrow \td{\Vh e_1}{\ell} &= -\left(\Vh e_1 \cdot \td{\Vh t}{\ell}\right) \Vh t.
\end{align}
where we have made use of the fact that $\td{\Vh e_n}{\ell} \cdot \Vh e_n = 0$ for any normal vector and $\Vh \theta = -\sin \theta\ \Vh e_1 + \cos \theta\ \Vh e_2$.
It can be shown that it automatically follows that $\td{\Vh e_2}{\ell} \cdot \Vh e_1 = 0$ provided they are part of an orthonormal set.

\Eqnref{de1-pt} defines the `parallel-transport' framing, also known as a Bishop or minimally rotating frame \cite{bishopThereMoreOne1975, wangComputationRotationMinimizing2008}.
This is the preferred framing for measuring twist since it eliminates framing contributions (i.e.~$\tau = \td{\theta_{PT}}{\ell}$).
It also has the desirable property that is easy to compute from numerical data and is relatively insensitive to noise because it can be computed without making use of higher order derivatives of the centerline \cite{wangComputationRotationMinimizing2008}.
Finally, a vortex bundle which is untwisted with respect to this frame has zero helicity dissipation (see \secref{helicity}).

By contrast, for the commonly used Frenet-Serret framing \cite{moffattHelicityCalugareanuInvariant1992} we obtain:
\begin{align}
    \Vh e_{1,FS} &= \Vh N = \frac{\td{\Vh t}{\ell}}{\left|\td{\Vh t}{\ell}\right|}\\
    \Vh e_{2, FS} &= \Vh B = \Vh t \times \Vh N\\
    \Rightarrow \tau &= \td{\theta_{FS}}{\ell} + \Vh N \cdot \td{\Vh B}{\ell}
\end{align}
where the last term is the negative of the centerline torsion, and $\theta_{FS}$ is the angle of a filament measured with respect to the Frenet-Serret framing.
Thus a bundle which is untwisted with respect to a Frenet-Serret framing would not in general conserve helicity.

\bibliography{From_Dye_to_Dynamics.bib}

\end{document}